\documentclass[epsfig,graphics,twocolumn,floatfix,a4paper,showpacs]{revtex4}
\usepackage{amsmath,amsfonts,amssymb,graphicx,color,times,bbm,psfrag,dsfont}
\usepackage[english]{babel}
\usepackage[latin1]{inputenc}
\usepackage[T1]{fontenc}

\newcommand{\bed}{\[}
\newcommand{\eed}{\]}
\newcommand{\beq}{\begin{equation}}
\newcommand{\eeq}{\end{equation}}
\newcommand{\beqa}{\begin{eqnarray}}
\newcommand{\eeqa}{\end{eqnarray}}
\newcommand{\ket} [1] {\vert #1 \rangle}
\newcommand{\bra} [1] {\langle #1 \vert}

\newcommand{\proj}[1]{\vert{#1} \rangle \langle {#1} \vert}

\newcommand{\gras}[1]{\bold{#1}}

\newcommand{\tr}{\mathop{\mathrm{tr}}}

\usepackage{graphicx}
\usepackage[mathscr]{eucal}
\usepackage{amsmath}
\usepackage{amssymb}
\usepackage{epstopdf}
\usepackage{epsfig}
\def\one{\ensuremath{\hbox{$\mathrm I$\kern-.6em$\mathrm 1$}}}
\def\tr{ \mbox{tr}}

\begin{document}

\title{
\vskip -50pt
\begin{small}
\hfill UB-ECM-PF 08/12    \\
\vskip  5pt
\end{small}
Scaling law for topologically ordered systems at finite temperature}

\author{S. Iblisdir$^{1}$, D. P\'erez-Garc\'ia$^{2}$, M. Aguado$^{3}$ and J. Pachos$^{4}$}

\affiliation{
 $^1$Dpt. Estructura i Constituents de la Materia, Universitat Barcelona, 08028 Barcelona, Spain\\
$^2$Dpt. An\'alisis Matem\'atico, Universitad Complutense de Madrid, 28040 Madrid, Spain\\
$^3$ Max Planck Institut f\"ur Quantenoptik, Max-Kopfermann-Str, Garching D-85748, Germany\\
$^4$ School of Physics and Astronomy, University of Leeds, Leeds LS2 9JT, United Kingdom}

\date{\today}

\begin{abstract}
Understanding the behaviour of topologically ordered lattice systems at finite temperature is a way of assessing their potential as fault-tolerant quantum memories.  We compute the natural extension of the topological entanglement entropy for $T > 0$, namely the subleading correction $I_{ \textrm{topo} }$ to the area law for mutual information.  Its dependence on $T$ can be written, for Abelian Kitaev models, in terms of information-theoretic functions and readily identifiable scaling behaviour, from which the interplay between volume, temperature, and topological order, can be read. These arguments are extended to non-Abelian quantum double models, and numerical results are given for the $D(S_3)$ model, showing qualitative agreement with the Abelian case.
\end{abstract}
\maketitle

The notion of topological order was first introduced in the context of the fractional quantum Hall effect \cite{wen90}. It aims at identifying phases that cannot be separated by  local order parameters. Such phases can exhibit exotic phenomena such as topologically protected ground state degeneracy or quasiparticle excitations, called anyons, with statistical behavior that is different from bosons or fermions \cite{wenbook}. Besides the fact that they reveal unusual states of matter, topologically ordered systems are also interesting because they might allow for intrinsically fault-tolerant quantum computation \cite{Kit97,review}. In such systems, the division of a quantum algorithm as initialisation, unitary evolution and read-out \cite{Preskill} would translate into creating anyons,  braiding them, and fusing them back to the vacuum respectively. While conceptually appealing, the robustness of this \emph{topological} quantum computation against realistic noise models is far from being fully assessed.

This paper is devoted to quantifying how temperature affects a topologically ordered medium. For that, we will use an entropic topological order parameter, $I_{\textrm{topo}}$, and focus on lattice spin systems that are exactly solvable \cite{Kit97} \footnote{Exactly solvable in the sense that their low energy sectors can be analytically worked out and are well understood.}. We will show that, at any fixed temperature, $I_{\textrm{topo}}$ is nonzero only if the size of the system is finite. Importantly, we will exhibit a \emph{scaling} relation describing how a given increase of the system size can be compensated by a vanishing decrease of the temperature. After recalling some notions on the topological entropy, we will provide a complete analysis for the simple toric code model. Then, we will turn to more general models, show how to compute entropic quantities, and provide numerical evidence for scaling laws in the case of the simplest non-abelian quantum double model, the one based on $D(S_3)$ \cite{Kit97}. We believe that our findings are relevant to topological quantum computation \cite{review}.
 
\begin{figure}[h]
\begin{center}
\includegraphics[width=25mm,height=25mm]{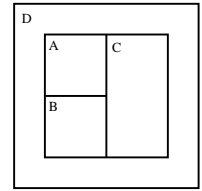} 
\caption{Division of a torus (or a sphere) into four regions.}\label{fig:torusdivided}
\end{center}
\end{figure}

We start our discussion by briefly recalling the notion of topological entropy \cite{Levin-Wen05:topent,Kit-Pre05}. It will play a central role in the following. Let us consider a two-dimensional many-body quantum system in a pure state of its ground subspace and let $R:R_c$ denote a bipartition of this system, with $R$ being connected. Let us further assume that our system satisfies an 'area' law. That is, the von Neumann entropy of region $R$ grows linearly with the size of its boundary, $S_R= -\tr \rho_R \ln \rho_R= \alpha' |\partial R|- \gamma'+\delta(R)$, where $\rho_R$ denotes the reduced density matrix corresponding to region $R$, $\alpha'$ is a constant, and where $\lim_{|\partial R| \to \infty} \delta(R)=0$. As first pointed out in \cite{Ham04}, and further analysed in \cite{Levin-Wen05:topent,Kit-Pre05}, a non-zero value for the constant correction $\gamma'$ reveals topological order. Indeed, $\gamma'$ is a topological invariant of the system and is related to the quantum dimension $\mathcal{D}$ of the model at hand as $\gamma'=\ln \mathcal{D}$. 

Let us consider a system defined on a torus or on a sphere, and divide it into four regions $A,B,C,D$ (see Fig.\ref{fig:torusdivided}). We will use 
\beq\label{eq:defparamorder}
I_{\textrm{topo}}=I_A+I_B+I_C-I_{AB}-I_{AC}-I_{BC}+I_{ABC}.
\eeq
as a topological order parameter. ($I_R=S_R+S_{R^c}-S_{R \cup R^c}$ denotes the mutual information between a region $R$ and its complement $R^c$.) This definition of $I_{\textrm{topo}}$ amounts to replace von Neumann entropies by mutual informations in the definition of topological entropy introduced in \cite{Kit-Pre05}. At zero temperature, $I_{\textrm{topo}}=-2 \gamma'$. Our choice is motivated by the fact that, at finite temperature, the mutual information between a region $R$ is a measure of the correlations of this region with its environment. The lattice systems we are going to study obey an area law:
\beq\label{eq:handy1}
I_R=\alpha |\partial R|-\gamma,
\eeq
and have a finite correlation length $\xi$  \cite{wolf}. Therefore, $I_{\textrm{topo}}$ is a quantity where  correlations due to a finite $\xi$ cancel out, and that reveals correlations due to topological order only, as the topological entropy, $\gamma'$, at zero temperature \cite{Levin-Wen05:topent}. Direct substitution of Eq. (\ref{eq:handy1}) into Eq.(\ref{eq:defparamorder}) shows that $I_{\textrm{topo}}=\gamma$ \footnote{The use of mutual information to study topological order in classical system has been discussed in \cite{topoclassic}.}.

We now want to compute $I_{\textrm{topo}}$ at finite temperature for the toric code \cite{Kit97}. Consider a torus $T$ tiled into $L \times L$ square plaquettes. Let us associate a two-level system ('spin') with each edge of this tiling and let us assume that these spins interact through the hamiltonian
\beq\label{eq:htoric}
H=-\sum_p B_p-\sum_s A_s,
\eeq
where the index $p$ (resp. $s$) runs over all plaquettes (resp. vertices) of the tiling. The operators $B_p$ and $A_s$ are defined as $B_p= \prod_{i \in p} \sigma_i^z$ and $A_s=\prod_{i \in s} \sigma_i^x$. The hamiltonian $H$ is made of local terms, all commuting with each other and, as it turns out, its eigenstates can have arbitrary values for the operators $B_p$ and $A_s$, up to the constraint $\prod_p B_p=\prod_s A_s=1$. Therefore, these eigenstates are labelled by a triple of quantum numbers: $\ket{\phi,c,w}$. A pattern $\phi$ denotes the position of all plaquette or ``flux-type'' excitations, a pattern $c$ indicates the position of all vertex or ``charge-type'' excitations, and $w$ indexes the topological degeneracy of the state for a fixed configuration of defects. The latter quantum number is made of two bits ($w_1$ and $w_2$) that label the values of the integrals of motion of $z$-operators around non-contractible loops on the torus (Wilson loops) \cite{Kit97} . We have
\beq
H \ket{\phi,c,w}= (E_0+2 |\phi|+2 |c|) \; \ket{\phi,c,w},
\eeq
where $E_0=-2L^2$ is the ground state energy, and $|\phi|$ (resp. $|c|$) denotes the number of flux excitations (resp. charge excitations) of the pattern $\phi$ (resp. $c$). If $P_i$ denotes the projector onto the sector of energy $E_i=E_0+4 i$ and $d_i=\tr P_i$ denotes its  dimension, we have that $d_i=4 \sum_{n_{\phi},n_{c} \leq L^2/2} \; \sum_{n_{\phi}+n_{c}=i} \binom{L^2}{2 n_{\phi}} 
\binom{L^2}{2 n_{c}}$.

We now compute all the thermodynamical quantities we need, assuming that our system is at thermal equilibrium at inverse temperature $\beta$. Some details of our calculations will be presented elsewhere \cite{ibli}. The partition function reads $Z(\beta,L)=\tr e^{-\beta H}=\sum_{i=0}^{L^2} e^{-\beta E_i} d_i=((2 \cosh \beta)^{L^2}+(2 \sinh \beta)^{L^2})^2$, and the von Neumann entropy of the state $\rho=e^{-\beta H}/Z(\beta,L)$ of the whole torus is derived from the identity
\beq\label{eq:stot}
S_{\textrm{tot}}=\ln Z-(\beta/Z) \partial Z/\partial\beta.
\eeq
In order to compute the von Neumann entropy of a region $R$, and eventually $I_{\textrm{topo}}$, we first observe that $\rho_R(\phi,c,w)=\tr_{R_c} \ket{\phi,c,w}\bra{\phi,c,w}$ ($R_c=T \backslash R$) depends on $w$ only if $R$ contains non-contractible loops ($w$ cannot be measured locally). Also, up
to total charge conservation, $\rho_R(\phi,c,w)$ will only depend on the excitations located on plaquettes and crosses inside $R$. Therefore, we distinguish between three types of flux excitations: those located on plaquettes with support on $R$, $\phi_R$, those located on plaquettes with support on $R_c$, $\phi_{R_c}$, and those located on plaquettes with support partially on $R$ and partially on $R_c$, $\phi_{\partial R}$. Similarly, we divide the charge excitations into $c_R, c_{R_c}, c_{\partial R}$. The  crucial observation that helps to compute $\rho_R=\tr_{R_c} e^{-\beta H}/Z(\beta,L)$ and eventually $S(\rho_R)$ is that $\partial R$-fluxes and $\partial R$-charges can be driven inside $R_c$ by application of Pauli operators acting on links of $R_c$, that is: $\ket{\phi,c,w}$ is of the form $U_{R_c} \ket{\phi_R,\phi_{R_c},c_{R},c_{R_c},w}$,
for some unitary operator $U_{R_c}$. To lighten the notations, let us denote by $\gras{q}_R$ and $\gras{q}_{R_c}$ the patterns of excitations (both flux and charge) in $R$ and in ${R_c}$ respectively. All excitations in $R$ (resp. ${R_c}$) can be fused into a single excitation $q_R^1$ (resp. $q_{R_c}^1$) through a unitary operator $U'_R(\gras{q}_R,q_R^1) \otimes U'_{R_c}(\gras{q}_{R_c},q_{R_c}^1)$ that relates $\ket{\gras{q}_R,\gras{q}_{R_c},w}$ and a state $\ket{q_R^1,q_{R_c}^1,w}$, such that $R$ and $R_c$ contain at most one excited plaquette (vertex). As in \cite{Kit97}, we will call \emph{site} a plaquette and an adjacent vertex. The possibly excited plaquette and the possibly excited vertex within $R$ (resp. $R_c$) can always be chosen to form a site. The state $\ket{q_R^1,q_{R_c}^1,w}$ can be obtained, from a ground state $\ket{\xi,w}$, by a product of unitary operators along a string connecting the site where $q_R^1$ is located to the site where $q_{R_c}^1$ is located. In summary, $\ket{\gras{q}w}$ can be written as $U_R(\gras{q}_R) \otimes U_{R_c}(\gras{q}_{R_c},\gras{q}_{\partial R}) \ket{\xi,w}$ for some unitary operators $U_R(\gras{q}_R)$ and $U_{R_c}(\gras{q}_{R_c},\gras{q}_{\partial R})$. ($\gras{q}_{\partial R}$ refer to the charges which are neither fully in $R$, nor fully in $R_c$.)

We are now in a position to compute $S_R$. The thermal state of the toric code reads
\beq
\rho=\sum_{w,\gras{q}} \frac{e^{-\beta(E_0+\Delta E |\gras{q}|)}}{Z(\beta,L)}
\ket{\gras{q},w}\bra{\gras{q},w},
\eeq
where $\Delta E=2$ is the energy associated with a single excitation, and where $\gras{q}=\gras{q}_R \cup \gras{q}_{R_c} \cup \gras{q}_{\partial R}$. Therefore the reduced density matrix of the system $R$ reads $\rho_R=\sum_{w}\sum_{\gras{q}_R,\gras{q}_{R_c},\gras{q}_{\partial R}}$ $\frac{e^{-\beta(E_0+\Delta E |\gras{q}|)}}{Z(\beta,L)}\tr_{R_c} [U_R(\gras{q}_R) \ket{\xi,w}\bra{\xi,w} U_R(\gras{q}_R)^{\dagger}]$. To simplify further this expression, we observe that two reduced states $\rho_R(\phi,c,w)$ and $\rho_R(\phi',c',w')$ are orthogonal whenever $(\phi_R,c_R) \neq (\phi'_R,c'_R)$. Indeed two such states can be perfectly distinguished through a measurement of $A_s$ or $B_p$ operators having support on $R$. Therefore, the sum $\sum_{\gras{q}_R}$ is actually a direct sum. If we assume that $R$ is contractible in neither of both directions on the torus, then the values of the Wilson loops $w$ can be revealed by measurements having support fully on $R$ and the sum $\sum_w$ also turns to be a direct sum. So, $\rho_R$ can be written as
\beq\label{eq:decrhoR}
\rho_R=\bigoplus_{w,\gras{q}_R} C(\gras{q}_R) \tr_{R_c} U_R(\gras{q}_R,x) \proj{\xi,w} 
U^{\dagger}_R(\gras{q}_R,x),
\eeq
where $4 C(\gras{q}_R)=\frac{4 e^{-2 \beta |\gras{q}_R|)}}{Z(\beta,L)} \sum_{\gras{q}_{\partial R},\gras{q}_{R_c}} e^{-\beta (E_0+2 |\gras{q}_{\partial R}|+2 |\gras{q}_{R_c}|)}$ is the marginal probability of a configuration of defects $\gras{q}_R$. It is the decomposition (\ref{eq:decrhoR}) that allows to compute $S_R$ exactly. From it, we find that $S_R$ separates into a ground state area contribution and a finite $\beta$ contribution
\beq\label{eq:strucS}
S_R= S_R^{\textrm{gs}}+V(\beta,N_p(R),L)+V(\beta,N_*(R),L),
\eeq
where $S_R^{\textrm{gs}}=S(\tr_{R_c} \ket{\xi,w} \bra{\xi,w})=(|\partial R|-1) \ln2$ \cite{Ham04}, and where $N_p(R)$ ($N_*(R)$) denotes the number of plaquettes (crosses) fully contained in $R$. The function $V$ can be computed exactly using elementary combinatorial identities \cite{ibli}. When $\rho_R$ is fully contractible, the direct sum over $w$ appearing in (\ref{eq:decrhoR}) should be replaced by a simple sum. As a result, the expression for $S_R$ picks a $-\ln4$ additive correction. 

We have used Eq.(\ref{eq:strucS}) to compute $I_{\textrm{topo}}$. We have found that at any finite $\beta$, $I_{\textrm{topo}}$ vanishes in the limit where $L \to \infty$, a result that is consistent with those of 
Ref.\cite{Cast06}, and indicates that (i) the toric code exhibits no temperature-driven phase transition for $I_{\textrm{topo}}$, and that (ii) the toric code is likely not a scalable quantum memory in the strictest sense \cite{Preskill} as far as temperature-induced errors are considered \footnote{The situation could be different in more than two dimensions \cite{dennis}.}. Our computations also reveal that the mutual information between a region $R$ and its complement satisfies an area law of the form Eq.(\ref{eq:handy1}). Computing $I_{\textrm{topo}}$ from it in the limit $L \to \infty$, and considering $|\partial R|=4 \nu L$, with $\nu < 1$, we have found a remarkably simple expression:
\beq\label{eq:largeL}
I_{\textrm{topo}}=2 \ln2 -h_2(p_p^{\textrm{even}})-h_2(p_*^{\textrm{even}}),
\eeq
where $h_2(x)=-x \ln(x)-(1-x) \ln(1-x)$ denotes the Shannon entropy of a binary outcome random variables, $p_p^{\textrm{even}} \simeq p_*^{\textrm{even}} \simeq (1+\theta^{\nu^2 L^2})(1+\theta^{(1-\nu^2) L^2})/2 (1+\theta^{L^2})$ is the probability that the region $R$ contains an even number of excited plaquettes (crosses), with $\theta=\tanh\beta$. Eq.(\ref{eq:largeL}) allows to understand simply why $I_{\textrm{topo}}$ vanishes at finite temperature when $L \to \infty$: in this limit, $p_p^{\textrm{even}} \simeq p_*^{\textrm{even}} \simeq 1/2$. This equation also reveals a scaling law: at fixed value of $\nu$, $I_{\textrm{topo}}$ only depends on $\beta$ and $L$ through the parameter $t=\theta^{L^2}$. In particular, a fixed value of $t$ (and thus a fixed value of the topological mutual information),  corresponds to the following relations between size and temperature:
\beq\label{eq:scaling1}
\beta(t,L)=\ln L-\frac{1}{2} \ln(\frac{1}{2} \ln \frac{1}{t})+O(L^{-2}),
\eeq 
\beq\label{eq:scaling2}
\frac{\partial T(t,L)}{\partial L}=\frac{-1}{L (\ln L-\frac{1}{2} \ln(\frac{1}{2} \ln \frac{1}{t})+O(L^{-2}))^2}+O(L^{-3}).
\eeq
This last relation tells us how an increase of the size of the system should be compensated by a decrease of temperature in order to maintain a fixed value of the topological entropy. We understand it as an important nuance over the fact that $I_{\textrm{topo}}$ vanishes when $L \to \infty$, at finite temperature. In particular, it shows that the \emph{rate} at which the temperature should be decreased, in order to maintain a fixed value of $\gamma$, \emph{decreases} with the size of the system. 

We now turn to a family of non-abelian models, those based on a quantum double. From now on, we will consider systems defined on an oriented lattice with the geometry of a sphere. A quantum degree of freedom with basis states labelled by the elements of a finite group $G$ is associated with each link of this lattice. These links interact through a hamiltonian of the form (\ref{eq:htoric}), with vertex opertators $A_s$ and plaquette operators $B_p$ that still commute. One could write down their form explicitly \cite{Kit97}, but it will not be necessary here. A natural way to deal with such models would be to start by providing a description of the complete set of eigenstates of $H$ similar to the one we have used for the toric code.  However, when $G$ is non-abelian, diagonalising $H$ seems to be a difficult problem. Yet, we can argue that we actually do not need to as far as we are only interested in the topological mutual information. The elementary excitations of $H$ live on sites (a site is a combination of a vertex and an adjacent plaquette). We will restrict to that part of the spectrum of $H$ such that excitations are elementary and pinned at fixed non-adjacent sites. This restriction can be thought of as additional error correction, where some plaquettes and vertices are over-protected so that they never get excited by thermal fluctuations (or only with vanishing probability). Therefore, we believe that  $I_{\textrm{topo}}$ for this modified model can only be larger than for the full spectrum. 

The space of $n$ excitations pinned at fixed sites has the structure \cite{Kit97}
\beq\label{eq:structhn}
\mathscr{H}[n]= \bigoplus_{q_1 \ldots q_n} \mathscr{H}_{q_1 \ldots q_n},
\eeq
where each index $q_i$ runs over all possible quasiparticle types for site $i$. Eq.(\ref{eq:structhn}) simply means that different excitation patterns lead to orthogonal states. Each space $\mathscr{H}_{q_1 \ldots q_n}$ splits as 
\bed
\mathscr{H}_{q_1 \ldots q_n}= \mathscr{K}_{q_1} \otimes \ldots \otimes \mathscr{K}_{q_n} \otimes 
\mathcal{M}_{q_1 \ldots q_n},
\eed
where the spaces $\mathscr{K}_{q_i}$ correspond to the local degrees of freedom of the quasiparticles
\cite{Kit97}. The fusion space, $\mathcal{M}_{q_1 \ldots q_n}$, is what makes non-abelian anyonic systems so special. For abelian models, for which the fusion rules are trivial, the dimension of the fusion space is equal to one. A non-trivial fusion space appears when fusing two anyons can yield different quasiparticles \cite{Preskill}:
\beq
q_a \times q_b= \sum_c N_{ab}^c \; q_c.
\eeq
The system has to fulfill some neutrality conditions, i.e. its state should be such that fusing all the particles yields the trivial particle, denoted $1$, with certainty \cite{bombin07}. The dimension of 
$\mathcal{M}_{q_1 \ldots q_n}$ depends on the tensor $N_{**}^*$ as follows \cite{Preskill}: $ \textrm{dim} \; \mathcal{M}_{q_1 \ldots q_n}= \sum_{b_1 \ldots b_{n-2}} N_{q_1 q_2}^{b_1} N_{b_1 q_3}^{b_2} \ldots  N_{b_{n-2} q_n}^1$.  We observe that computing this quantity amounts to contracting a (quasi) translationally invariant matrix product state \cite{mps}. Likewise $\textrm{dim} \mathscr{H}[n]$ can be computed efficiently, as well as the partition function of the model 
\bed
Z(\beta,n)=\sum_{q_1} \ldots \sum_{q_{n}} \prod_{i=1}^n d(q_i) 
\textrm{dim} \; \mathcal{M}_{q_1 \ldots q_n} e^{-\beta E(q_1 \ldots q_{n})}, 
\eed
where $E(q_1 \ldots q_{n})$ is the energy associated with a defect configuration $q_1 \ldots q_{n}$, and where $d(q_i)$ is the dimension of the space $\mathscr{K}_{q_i}$.

To compute $I_{\textrm{topo}}$, we consider a situation where a pair of anyons $\ket{q \bar{q}}$ is created in such a way that anyon $q$ lies in some region $R$ and its anti-quasiparticle $\bar{q}$ lies in the complementary region. The von Neumann entropy of region $R$ then reads 
\cite{Kit-Pre05}
\beq
S(\rho_R)=S(\rho_R^{\textrm{g.s.}})+\ln d_q.
\eeq

The entropy of a region when the system is in a thermal state can be computed once we are able to calculate the entropy of a region when the system is an arbitrary defect configuration. In turn, just as for the toric code, the latter entropy reduces to computing the entropy when there are only \emph{two} anyons in the system, and one lies inside the region we are interested in. Let $n_R$ and $n_{R_c}$ denote the number of sites contained in region $R$ and ${R_c}$ respectively ($n_R+n_{R_c} \equiv n$). Let $\gras{q}_R=q_1 \ldots q_{n_R}$ label the configurations of types of anyons living on the sites contained in region $R$. The total fusion space splits as:

\beq\label{eq:decompspace}
\mathcal{M}_{\gras{q}_R,\gras{q}'_{R_c}}=
\bigoplus_b \mathcal{M}^b_{\gras{q}_R} \otimes \mathcal{M}^{\bar{b}}_{\gras{q}'_{R_c}}
\otimes \mathcal{M}^1_{b,\bar{b}},
\eeq
where $\mathcal{M}^b_{\gras{q}_R}$ denotes the fusion space associated with event where all $\gras{q}_R$ quasiparticles fuse to an anyon of type $b$.

Due to total anyonic charge conservation, $\textrm{dim}\mathcal{M}^1_{b,\bar{b}}=1$. The decomposition (\ref{eq:decompspace}) induces the following representation of the thermal state of $H$:

\bed
\rho_{\textrm{th}}=\bigoplus_{\gras{q}_R,\gras{q}'_{R_c},b,\mu_1,\mu_2,s_R,s_{R_c}} 
\frac{e^{-\beta (E(\gras{q}_R)+E(\gras{q}'_{R_c}))}}{Z(\beta,n)} 
\eed
\beq
\mathscr{P}(\gras{q}_R \to b,\mu_1, s_R; \gras{q}'_{R_c} \to \bar{b},\mu_2,s_{R_c}),
\eeq
where the projector  $\mathscr{P}(\gras{q}_R \to b,\mu_1, s_R; \gras{q}'_{R_c} \to \bar{b},\mu_2,s_{R_c})$ refers to a pure state where all $\gras{q}_R$ (resp.$\gras{q}'_{R_c}$) quasiparticles contained in $R$ (resp. $R_c$) fuse to $b$ (resp. $\bar{b}$) through the channel $\mu_1, \;  1 \leq \mu_1 \leq \textrm{dim} \mathcal{M}^b_{\gras{q}_R}$ (resp. $\mu_2, \;  1 \leq \mu_2 \leq \textrm{dim}\mathcal{M}^{\bar{b}}_{\gras{q}'_{R_c}}$). $E(\gras{q}_R)=\sum_{j=1}^{n_R} E(q_j)$ denotes the energy associated with the configuration $\gras{q}_R$, and $s_R$ is a collective index for the internal degrees of freedom of the quasiparticles contained in region $R$, $1 \leq s_R \leq d(\gras{q}_R)=\prod_{j=1}^{n_R} d(q_j)$. $E(\gras{q}'_{R_c})$ and $s_{R_c}$ are defined likewise for the region $R_c$. Note that the ground state energy is now made equal to zero, upon shifting the hamiltonian by a multiple of the identity \cite{Kit97}.

The von Neumann entropy of the reduced state of the region $R$ now reads 
\bed
S(\rho_R)=
\sum_b [\frac{Z_R(\beta,b) Z_{R_c}(\beta,\bar{b})}{Z(\beta,n)} (S^{\textrm{gs}}_R+\ln d_b
\eed
\beq\label{eq:entropyregionNA}
-\ln Z_{R_c}(\beta,\bar{b}))]
+\sum_b \frac{Y_R(\beta,b) Z_{R_c}(\beta,\bar{b})}{Z(\beta,n)}+ \ln Z(\beta,n),
\eeq
where $Z_R(\beta,b)=\sum_{\gras{q}_R} d(\gras{q}_R) e^{-\beta E(\gras{q}_R)} \textrm{dim} \mathcal{M}^{b}_{\gras{q}_R}$, and where $Y_R(\beta,b)=\beta \frac{\partial}{\partial \beta} Z_R(\beta,b)$. The von Neumann entropy of the whole sphere is given by Eq.(\ref{eq:stot}), while the ground state entropy reads \cite{ibli}: $S_R^{\textrm{gs}}=\ln|G| (|\partial R|-1)$. 

\begin{figure}[h]
\begin{center}
\includegraphics[width=60mm,height=35mm]{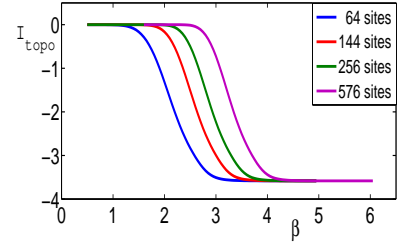} 
\caption{Topological quantum mutual information as a function of the inverse temperature $\beta$ for the $D(S_3)$ model. (Color online.)}\label{fig:nadata}
\end{center}
\end{figure}

Eq.(\ref{eq:entropyregionNA}) has allowed us to study numerically how the topological mutual information behaves as a function of $\beta$, for $G=S_3$, the smallest non-abelian finite group. Our results are shown on Fig.\ref{fig:nadata}. The systems we have considered are four tiled spheres, all with $96 \times 96$ plaquettes. The first sphere contains $n=64$ sites, the second $144$, the third $256$ and the fourth $576$. Although these systems are small, they are large enough to show that non-abelian systems are affected by temperature in the same way as the toric code: (i) For a fixed number of sites, there exists a finite width topological phase. (ii) $I_{\textrm{topo}}$ tends to $0$ for small values of $\beta$, and to $-3.58352=-2 \ln 6$ for sufficiently large values of $\beta$, as expected since the order of $S_3$ is six. (iii) The larger the number of sites, the larger the value of $\beta$ where the topological mutual information vanishes. Finally, we have observed that when $I_{\textrm{topo}}$ is plotted as a function of $n e^{-2\beta}$, the curves collapse when $n$ becomes large. We take this observation as strong evidence that the behaviour of $I_{\textrm{topo}}$ for the $D(S_3)$ model is governed by the \emph{same} scaling variable as the toric code. Indeed, $\ln(\theta^{L^2}) \simeq L^2 e^{-2\beta}$ in the limit where $\beta$ and $L$ are large ($n \propto L^2$ for the toric code). We therefore believe that the discussion held after Eq.(\ref{eq:largeL}) also holds in this case, and more generally for any quantum double model.

In conclusion, we have shown that the interplay of thermal effects, lattice size, and topological order (as measured in the subleading correction to the area law) is encoded in well-defined scaling relations (\ref{eq:scaling1})-(\ref{eq:scaling2}). In particular, the rate with which the temperature should be decreased to fight the effect of thermal fluctuations vanishes in the limit of large lattices. These relations seem to hold for both Abelian and non-Abelian systems. As a byproduct, we have derived a formula for the entropy of a region for non-abelian quantum double models defined on a lattice. This formula depends on the model only through the fusion tensor $N_{**}^*$ and through the energy associated with each quasiparticle. It is therefore tempting to believe that it holds in a more general context. Finally, it is appealing to try to give an operational meaning to $I_{\textrm{topo}}$, by connecting its value with the possibility of using a topologically ordered medium as a robust quantum memory. But to the best of our knowledge, such a  connection is still an open problem, even at zero temperature. Eq.(\ref{eq:largeL}) might contribute establishing it.

\emph{Acknowledgements.} It is a pleasure to thank G. K. Brennen, P. Calabrese, J.I. Cirac, J. I. Latorre and M. A. Martin-Delgado and A. Stern for fruitful discussions. Financial support from the Generalitat de Catalunya, MEC (Spain), QAP (EU), and Spanish grants MTM2005-00082, I-MATH, and CCG07-UCM/ESP-2797.

\end{document}